\documentclass[reprint, twocolumn]{revtex4}
\usepackage{amssymb}
\usepackage{graphicx}
\usepackage[colorlinks=true,citecolor=blue,linkcolor=black]{hyperref}

\begin{document}

\title{Orientated energy absorption from mid-infrared laser pulses in constrained water systems}
\author{Rong-Yao Yang}\email{ryyang@seu.edu.cn}
\author{Wei-Zhou Jiang}
\email{wzjiang@seu.edu.cn}
\author{Pei-Ying Huo}
\affiliation{School of Physics, Southeast University, Nanjing 211189, China}


\begin{abstract}
The energy acquisition based on resonant excitations are of great importance in chemical and biological systems. Here, the intramolecular resonant absorption of polarized mid-infrared pulses by bulk water and surface water is investigated using molecular dynamics simulation. The consequent heating based on the OH stretching vibrations is found to be very prompt, achieving more than 100 $K$ temperature jump under irradiation of a pulse with 1 $ps$ width and maximum intensity of 0.5 $V/nm$. A general anisotropic phenomenon is manifested as a result of preferential resonant excitation of symmetric or asymmetric OH stretching vibration, depending on the relationships between the  orientations of water molecules and the polarized direction of the pulse.
In the case of water molecules with the preferred dipole orientation, constrained by applied static electric field or spacial confinement, parallel to the polarized direction of the pulses, the energy absorption is dominated by the symmetric stretching mode (around 99 THz), while in the perpendicular case, the asymmetric stretching mode (around 101 THz) is more efficient.
Since orientated water molecules are prevalent in chemical and biological systems, these findings concerning orientation-dependent excitation of intramolecular vibrations are of special significance to understood the energy absorption and transition in relevant biochemical processes.
\end{abstract}

\maketitle




\section{Introduction}

Prompt and efficient energy acquisition from absorption of electromagnetic field plays an important role in aqueous environment. The resulting temperature jump could induce evolution in chemical reactions~\cite{MaWan-193,2019Cannelli289} and biological dynamics, such as DNA melting~\cite{MaWan-196} and protein unfolding \cite{MunozThompson-279,1998Dyer285,EbbinghausDhar-197}. The resonant excitations of the intramolecular or intermolecular vibrational modes pave a way to vast energy absorption in ultrashort time. It was found that explosive boiling of water can be achieved by direct vibrational excitation of water molecules with a infrared (IR) laser pulse~\cite{TakamizawaKajimoto-192}.
Lots of fascinating features could be aroused from vibrational excitation. For instance, a large enhancement of calcium ion permeation in an bio-channel was reported under the irradiation of a 42.5 $THz$ field which provokes the -COO$^-$ symmetric stretching mode~\cite{LiChang-230}, and mid-infrared (MIR) stimulation was manifested to have potential applications in neuronal signaling modulation~\cite{LiuQiao-233}. The resonances in water systems are of particular interest. There are abundant researches focusing on the electromagnetic field effects on water systems, such as transition of confined water to a superpermeation phase stimulated by 1.39 $THz$ field~\cite{ZhuChang-231}, enormous water flux amplification in nanotube triggered by tube vibration or 16 $THz$ electric field~\cite{ZhangJiang-162,ZhangYang-163}, and ultrafast ice melting caused by a 100 $THz$ laser pulse~\cite{Caleman-1}. Fast heating of water was investigated by resonant excitation of intermolecular vibrations with 1-30 $THz$ laser pulses~\cite{HuangYang-6,YangHuang-7,MishraBettaque-8}. However, a systematic research on the energy absorption of water by exciting the intramolecular vibrational modes is still absent.

Water molecule, as a non-linear molecule, has a large electric dipole moment that contributes eminently to the complex hydrogen-bond network. We have known that the fluctuating characteristic of this strong hydrogen bonding is a hotbed for many of the anomalous properties of water. For instance, a cascade of anomalies occurs in liquid water owing to the strong coupling of its orientational and translational order~\cite{ErringtonDebenedetti-28}.
A very bizarre water freezing happens on a pyroelectric surface during heating up process  resulting from turnover of the electric field direction~\cite{EhreLavert-205}. It was also reported that ice-liquid phase transition of spatially confined monolayer water can be triggered by a perpendicularly applied electric field~\cite{QiuGuo-40}. These marvelous phenomena suggest that properties of water molecules could exhibit direction-dependent features under external electric fields or dimensional confinements which make the hydrogen-bond network different from bulk water. Theoretically, the anisotropic facet in absorption of electromagnetic fields by constrained water systems usually can be roughly understood by the IR spectroscopy extracted from the dipole autocorrelation function based on equilibrium simulations.
On the other hand, the morphosis of water molecules could experience significant changes, apart from its equilibrium configuration, during the dynamic evolution in response to the strong external field exerted by a laser.
Considering the prevalence of the water and energy transports in biosystems and nano-devices in the electromagnetic bath~\cite{MaWan-193,ZhangYang-163}, we thus investigate here the energy absorption of water under the irradiation of ultrashort MIR pulses and the corresponding orientation-dependent features within it.
It will be shown that conditional resonant excitation of intramolecular vibrations in water leads to prompt and orientated heating under MIR pulses, which is meaningful for understanding the energy absorption processes in constrained water-involving systems.


\section{Computational methods}

\begin{figure}[htb]
\centering
\includegraphics[height=3.0cm,width=3.0cm]{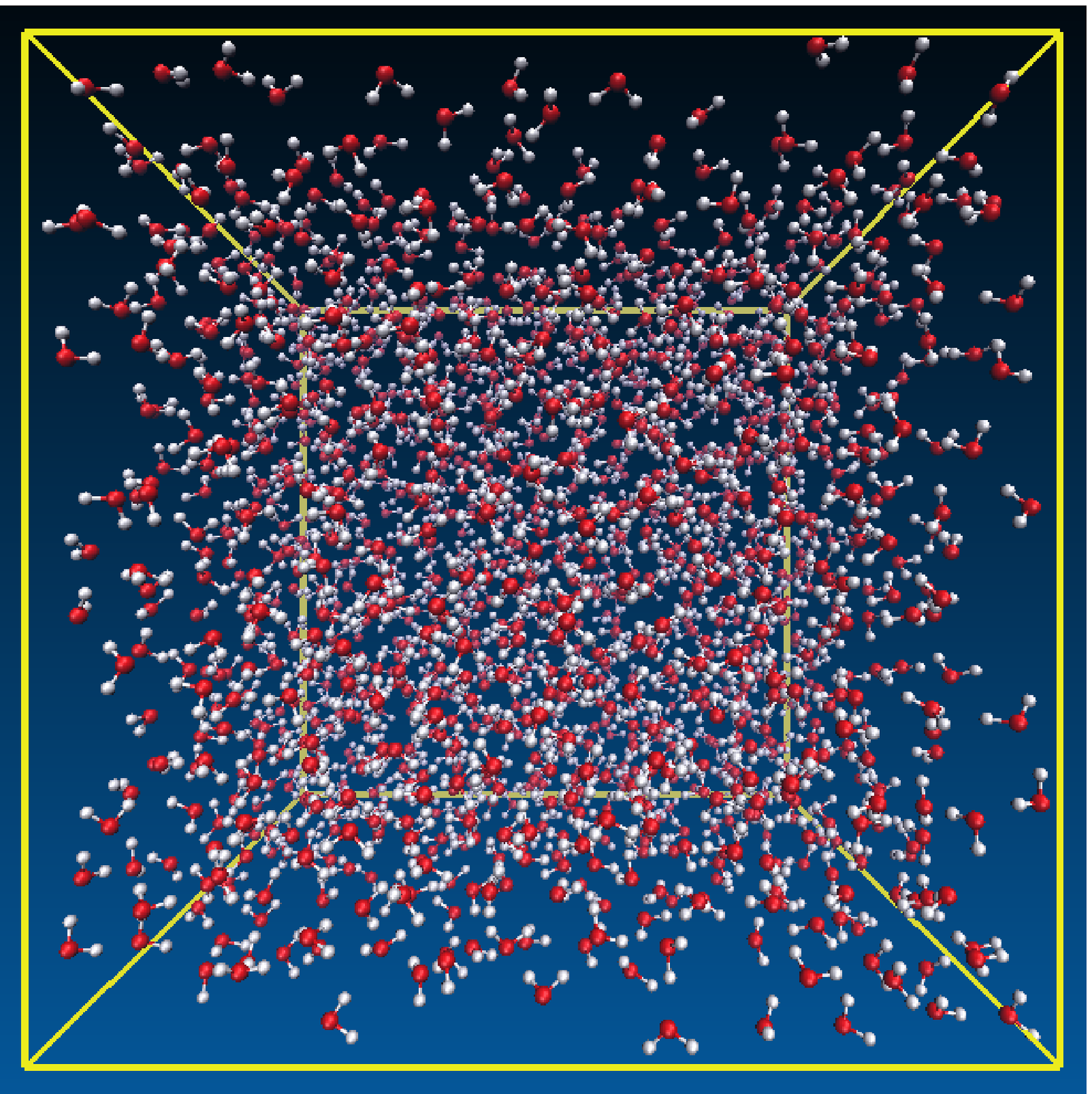} \includegraphics[height=3.0cm,width=3.0cm]{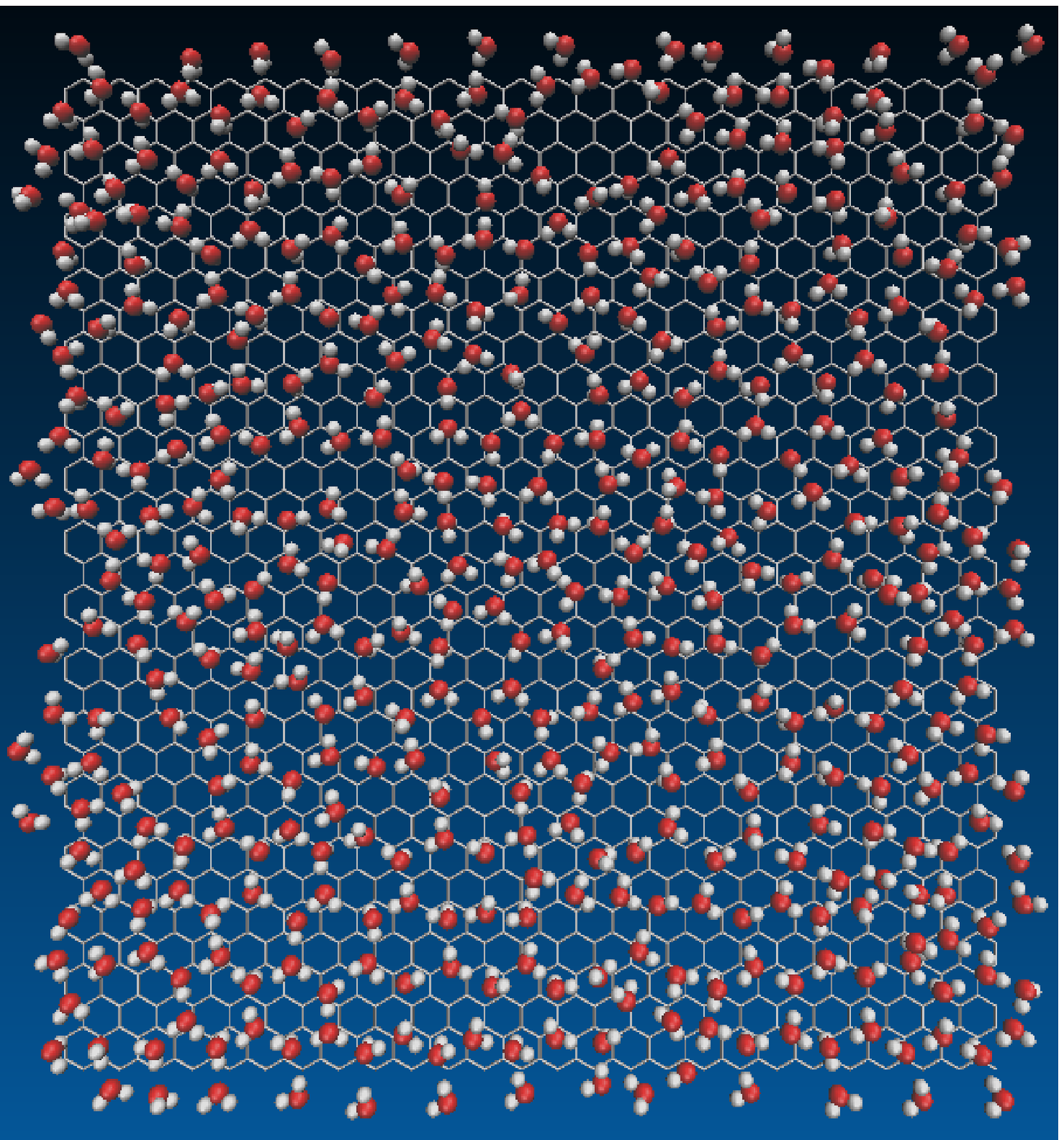}
\caption{(Color online) Snapshots of the investigated water systems created by VMD~\cite{HumphreyDalke-204}. The bulk water (left) is a $3.6\times3.6\times3.6 \ nm^3$ cubic box containing 1560 water molecules (density equals 1 $g/cm^3$), the surface water (right) is a monolayer of 384 water molecules on a $6.3\times6.4\ nm^2$ hexagonal surface (z axis perpendicular to the surface).
 }\label{F-models}
\end{figure}

In this work, we investigate energy in bulk water and surface water under the irradiation of linearly-polarized MIR laser pulses. The snapshots of the concerned systems are shown in Fig.~\ref{F-models}, including a periodic water box and a water monolayer on a surface of hexagonal lattice. The molecular dynamic simulation package NAMD2~\cite{PhillipsBraun-201} and the charmm force field~\cite{MackerellBashford-202} are adopted here. Periodic boundary conditions are applied to all systems. Flexible TIP3P water model is used where the bond interactions are described by harmonic potentials~\cite{JorgensenChandrasekhar-203}. A general process is as follow: an anterior 1 $ns$ equilibrium simulation under canonical assemble with Langevin dynamics gives the starting point ($t=0$) of a sequential $5\ ps$ long non-equilibrium simulation without temperature control under the irradiation of a MIR pulse. The time step in the simulation is 0.2 $fs$ and the initial temperature is 300 $K$. The amplitude of the electric field of the laser pulses has the form of $E=Acos(2\pi f t)e^{(t-t_c)^2/2\sigma^2}$. Unless otherwise mentioned, the maximum amplitude $A$ is $0.5\ V/nm$, the full width at half maximum (FWHM, equals $2.355\sigma$ ) is $1\ ps$ and pulses are centered at $t_c=2.5\ ps$. The interested pulse's frequency $f$ is in the range of 95-105 $THz$, i.e. in the domain of water's OH stretching vibrations. To reveal the anisotropic aspect of energy absorption, the MIR pulses are linearly polarized. For example, an electric field ${\bf E}=(0,\ 0,\ E)$ represents a pulse polarized along $z$ axis. The influence of the magnetic field of the pulse is neglected here since the Lorentz force in this case is orders of magnitude less than the electric force. The value of temperature jump induced by the MIR pulse is defined as the average temperature difference between the last picosecond ($t=4 \rightarrow 5\ ps$) and the first picosecond ($t=0 \rightarrow 1\ ps$) during the non-equilibrium simulations. Six independent simulations are conducted to minimize the statistical uncertainties. By scanning the frequency and changing the polarized direction of the pulses, we can obtain the frequency and orientation-dependent profiles of the energy-absorption efficiency.

To capture a first glimpse into the physical picture of orientated energy absorption, we also perform a simple test in polarized bulk water, where an additional static electric field (SEF) of 1 $V/nm$ along $z$ axis is applied to the water box in Fig.~\ref{F-models}. In this case, a preferential dipole orientation of water molecules along $z$ axis is achieved. There are similar systems in real world, such as bulk water influenced by a uniformly charged surface~\cite{EhreLavert-205}. By setting the pulse's polarization direction parallel or perpendicular to the preferential orientation of water dipoles, the directional-dependent energy absorption and heating effect could be examined. On the other hand, water configurations exhibiting preferential dipole orientation are prevalent in low dimensional systems, especially in biological or chemical systems. We thus explore the situations in two dimensional surface water on a hexagonal lattice irradiated by polarized MIR pulses with polarization directions parallel or perpendicular to the surface. This simulated prototypic surface is selected as the same as Ref.~\cite{Wang2009-68} with a 0.6$e$ charge modification. It is worth mentioning that water behaviors on this kind of surfaces have been experimentally verified in several real materials, see~\cite{WangYang-269} and references therein for details.
All the atoms composing the surface are fixed during the simulations. Based on these representative systems, the orientation-dependent energy absorption and heating under irradiation of MIR pulses can be manifested. 

\section{Results and discussion}

\begin{figure}[thb]
\centering
\includegraphics[height=4.5cm,width=6.0cm]{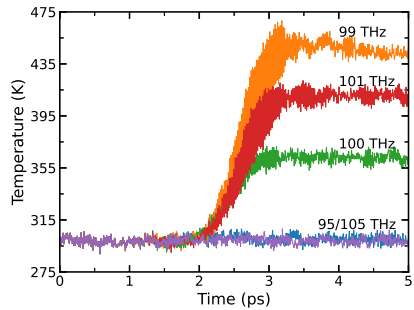}
\caption{(Color online) The temperature evolution with time for normal bulk water under irradiation of MIR pulses with several representative frequencies.
} \label{F-Texp}
\end{figure}

\begin{figure}[thb]
\centering
\includegraphics[height=4.0cm,width=6.0cm]{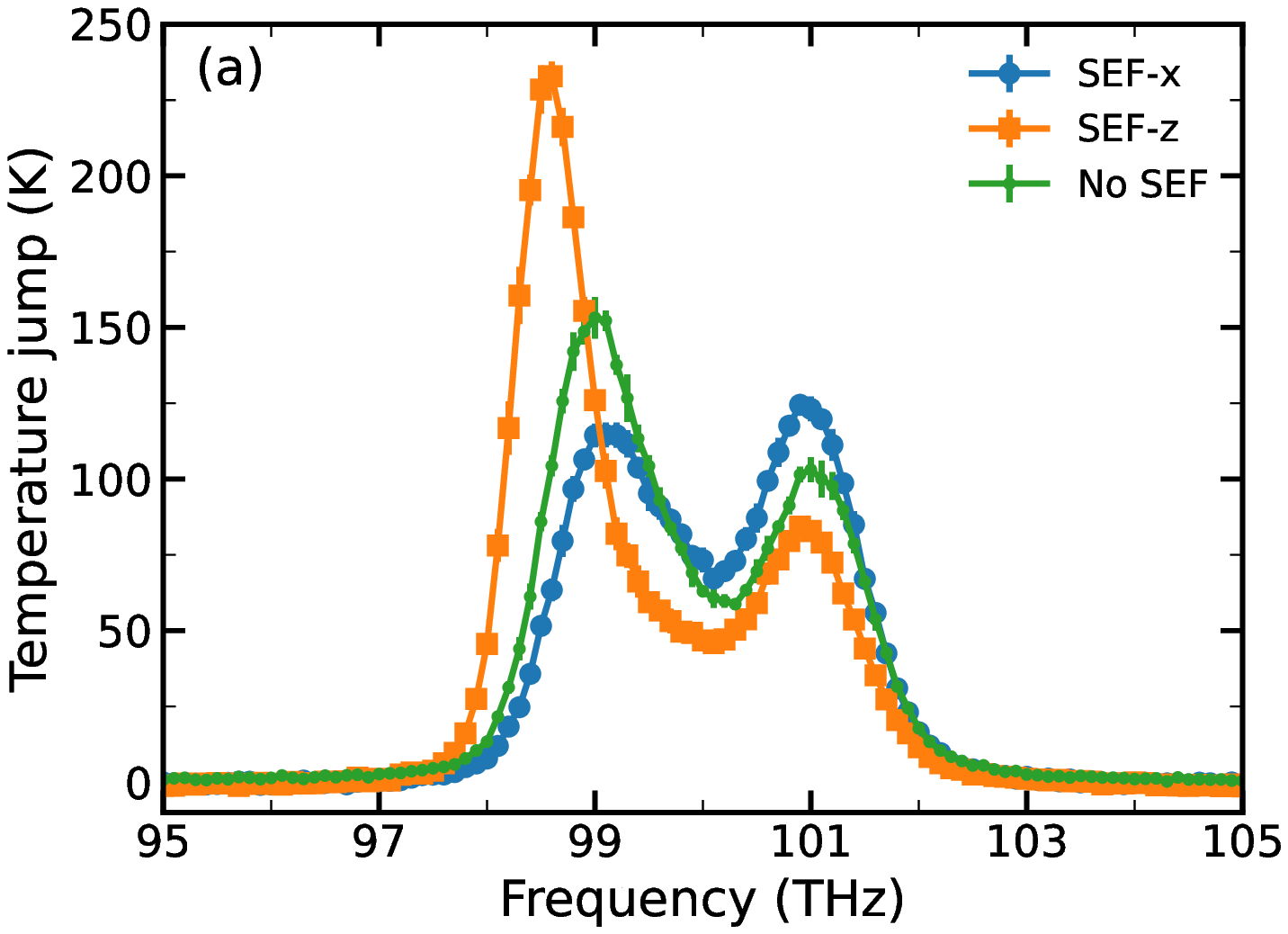}
\includegraphics[height=4.0cm,width=6.0cm]{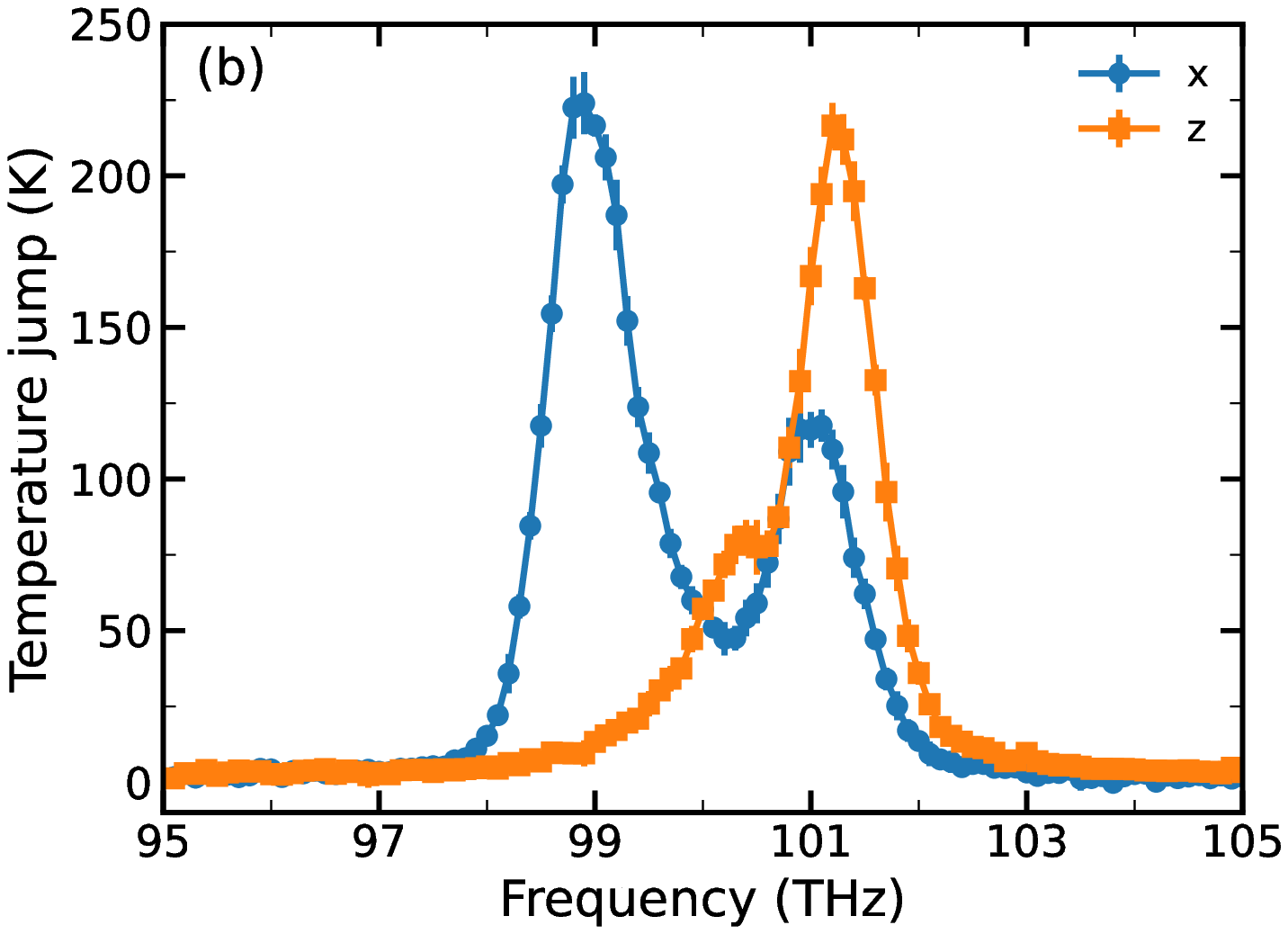}
\caption{(Color online) Temperature jump for various water systems after the irradiation of the MIR laser pulses with different frequencies. (a) is for bulk water, where the ``No SEF" denotes normal bulk water, the ``SEF-z" and ``SEF-x" represent results in polarized bulk water constrained by SEF along $z$ axis, (b) is for surface water. The ``x" and ``z" in the legends denote the polarization directions of the pulses. The error bars represent the standard deviations of six independent simulations, which are generally very small in these cases.} \label{F-T-Freq}
\end{figure}

At the beginning, we probe the heating effect caused by absorption of MIR pulses in normal bulk water. The pulses are linearly polarized along $z$ axis and changes in its direction will make no difference since water dipoles are uniformly distributed in all directions for unpolarized bulk water in the absence of external SEF. The typical patterns of temperature evolving with time for bulk water are shown in Fig.~\ref{F-Texp} for several representative frequencies. At frequency of 95 or 105 $THz$, the temperature just slightly fluctuates around the initial value 300 $K$ during the whole simulation time, suggesting no obvious heating effect exerted by the MIR pulses at these frequencies. However, the pulse at 99 or 101 $THz$ induces violent temperature jump ($>100\ K$) during the pulse's 1 $ps$ FWHM, while the case at 100 $THz$ shows a moderate temperature jump. These preliminary results indicate a characteristic of resonant excitations. The temperature jump in normal bulk water induced by the pulses with the frequency in range of 95-105 $THz$ is displayed in Fig.~\ref{F-T-Freq}a (labeled with ``No SEF").
Prompt heating occurs under the irradiation with pulse frequency in the range of 98-102 $THz$, which is in accordance with experimental results~\cite{BertieLan-261}. The maximum temperature jump, about 150 $K$, is achieved at 99 $THz$, which is much larger than the case in the intermolecular vibration region around 20 $THz$~\cite{HuangYang-6,YangHuang-7}. A fine structure in Fig.~\ref{F-T-Freq} can be spotted with doublet peaks at 99 and 101 $THz$, originating from resonant absorption of the laser pulses through excitations of the symmetric and asymmetric OH stretching modes respectively. The appearance of this obviously distinguished double peaks is actually a defect of the flexible TIP3P water model where the strong intra and inter molecular correlations are not adequately considered. This problem can be cured by using a flexible TIP4P model~\cite{LawrenceSkinner-264} or ab initio models~\cite{Heyden2010-12} at the cost of simulation time. However, TIP3P water model is good enough to clearly demonstrate the orientation-dependent absorption of laser pulses and the main conclusions herein are still reliable due to its generality in resonant processes.

\begin{figure}[thb]
\centering
\includegraphics[height=4.0cm,width=6.0cm]{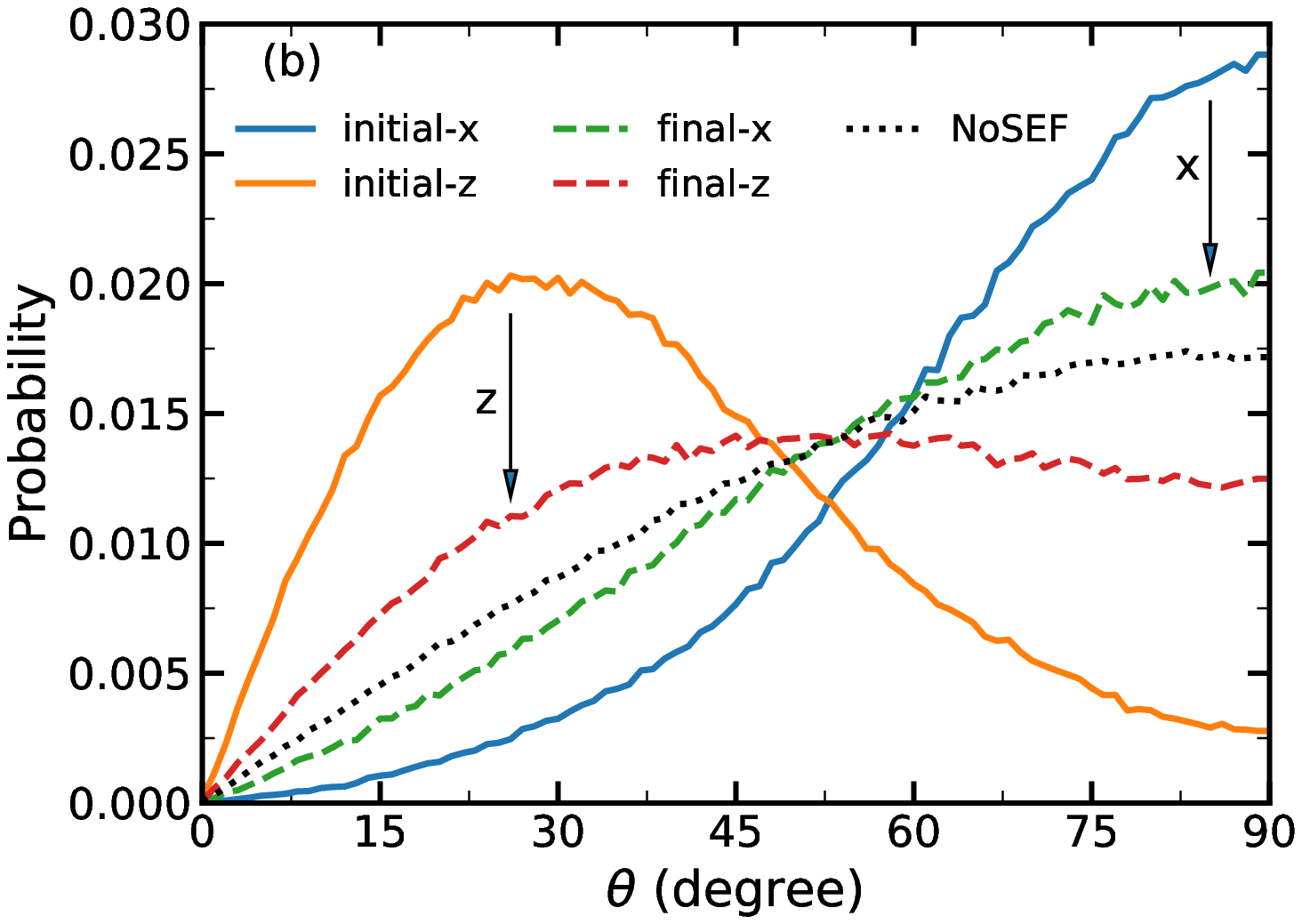}
\includegraphics[height=4.0cm,width=6.0cm]{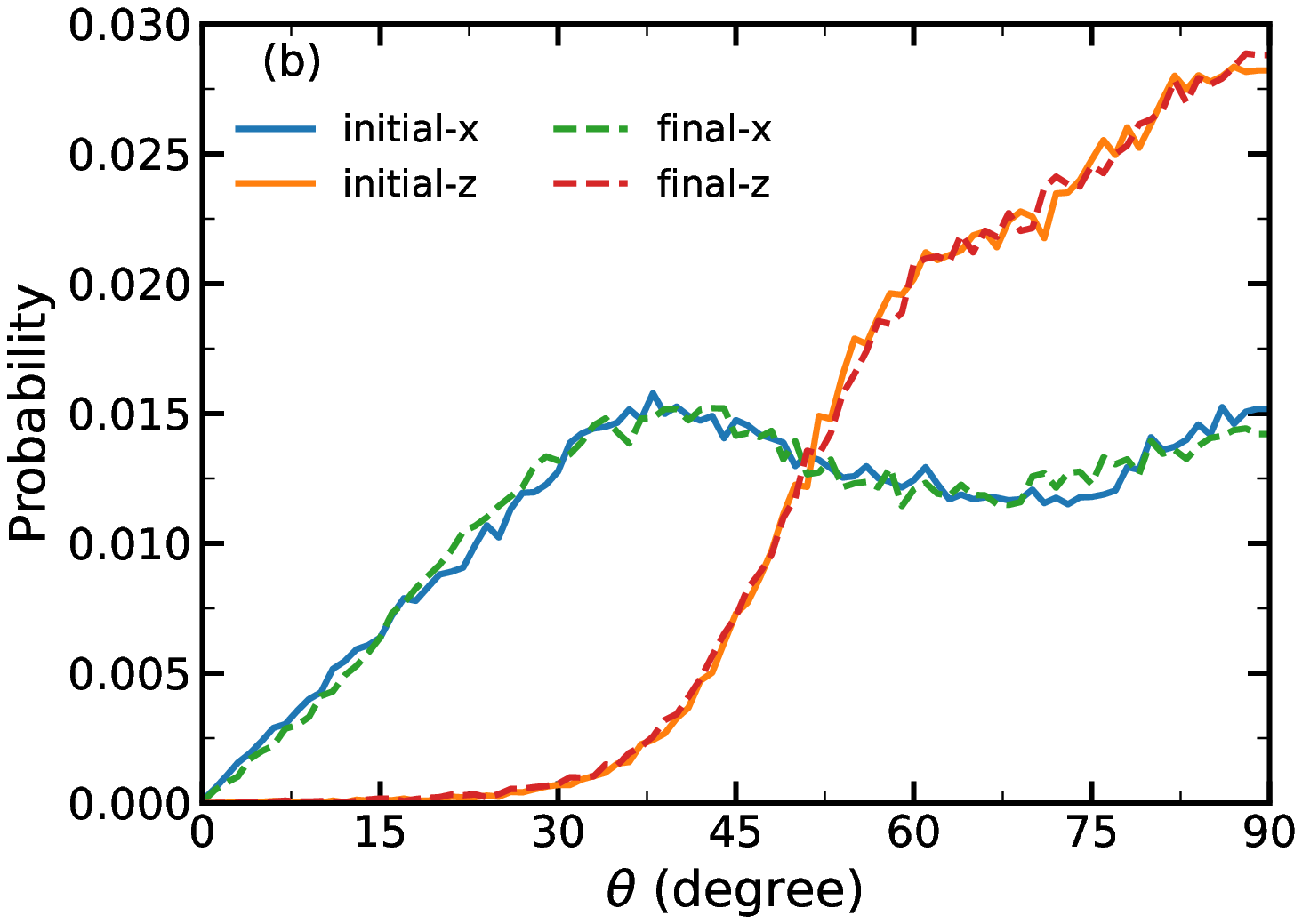}
\caption{(Color online) Probability distributions of angle between water dipoles and $x$ or $z$ axis in the polarized bulk water with SEF constraint (a) and surface water (b). The labels ``initial" and ``final" represent results counted at the first 1 $ps$ (initial state at 300 $K$) and the last 1 $ps$ (heated state) in the non-equilibrium simulation with a 99 $THz$ pulse, respectively. The dotted curve in (a) shows result for unpolarized bulk water whose integration always equals 1/3 for arbitrary directions, before or after pulses' irradiation.} \label{F-ang}
\end{figure}

We proceed to investigate the anisotropic heating effect in polarized bulk water constrained by an external SEF of 1 $V/nm$ along $z$ axis, which shall provide some conceptual information on orientation-dependent characteristic of water heating by MIR pulses. A net dipole moment will be produced as water molecules tend to align with this SEF, see Fig.~\ref{F-ang}a.
Shown in Fig.~\ref{F-T-Freq}a are the temperature jumps after the pulse radiation with polarization directions parallel or perpendicular to $z$ axis. Compared with results in normal bulk water, a prominent phenomenon in this polarized bulk water irradiated by pulses parallel to $z$ axis (labeled with ``SEF-z") is that the double peak structure becomes highly asymmetric with one dominant peak at 99 $THz$ and a largely weakened peak at 101 $THz$.
The result is inverse for the situation with pulses polarized along $x$ axis: the symmetric stretching peak at 99 $THz$ is reduced and the asymmetric stretching peak at 101 $THz$ is enhanced slightly. This seesaw of enhancement and weakening of heating efficiency stems from the alternative excitation of the symmetric and asymmetric OH stretching modes in orientation-preferential water under SEF constraint. Similar enhancement or reduction also appears around 53 $THz$ associating with the selective excitation of HOH bending vibration. These anisotropic phenomena demonstrate different energy absorption processes depending on water's orientational arrangements.


Abundant configurations of water exist in nature. Among them water systems in low dimension are of particular interest. As an example, a water monolayer on a hydrophilic surface of hexagonal lattice is adopted here to demonstrate the directional-dependent heating of water.
Fig.~\ref{F-T-Freq}b shows the temperature jump in the case of the pulses' polarization direction parallel or vertical to the surface. The seesaw of enhancement and weakening of heating efficiency due to the absorption of polarized laser pulses through symmetric and asymmetric OH stretching excitations becomes more noticeable, compared to results in the polarized bulk water. To realize a higher efficient energy absorption one should utilize pulses polarized vertical to the surface at symmetric OH stretching frequency near 99 $THz$ or parallel polarized pules at asymmetric OH stretching frequency near 101 $THz$. We mention that the corresponding results for water droplet on a hydrophobic graphene surface are qualitatively the same as that in normal bulk water. This pronounced anisotropic feature of surface water irradiated by MIR pulse illustrates again the importance of detailed water morphoses during the energy absorption dynamics in water-involved systems.

\begin{figure}[thb]
\centering
\includegraphics[height=4.0cm,width=6.0cm]{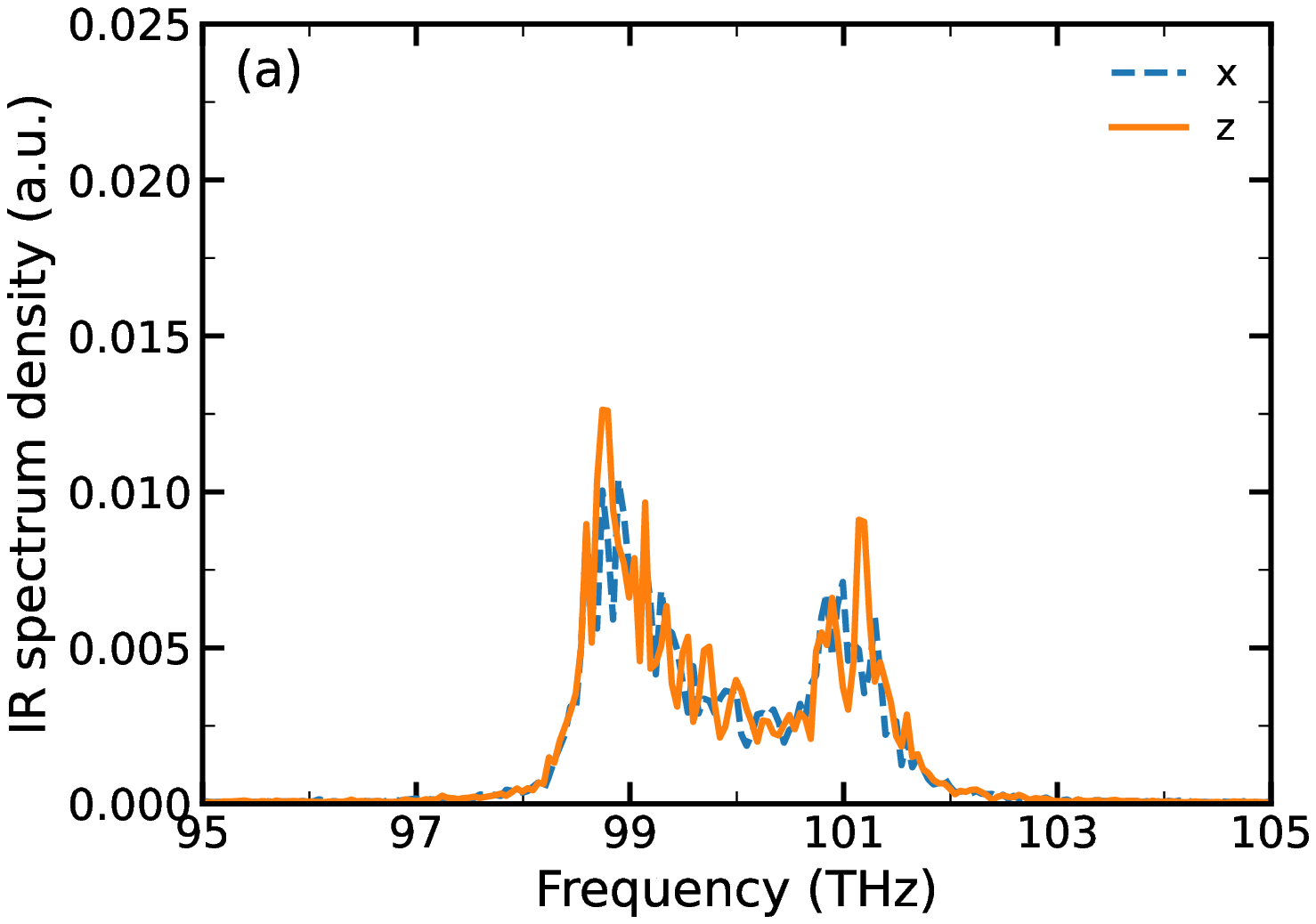}
\includegraphics[height=4.0cm,width=6.0cm]{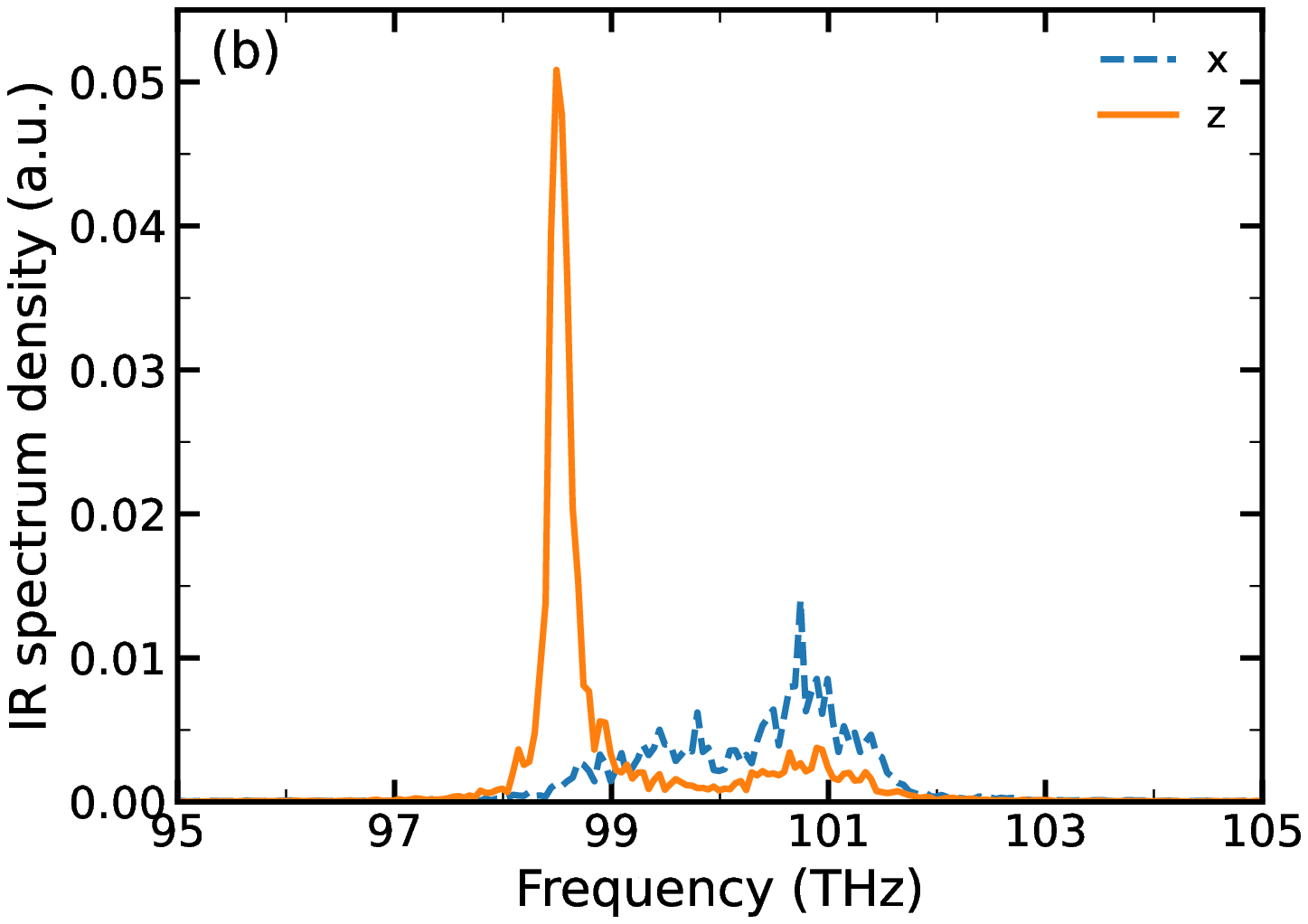}
\includegraphics[height=4.0cm,width=6.0cm]{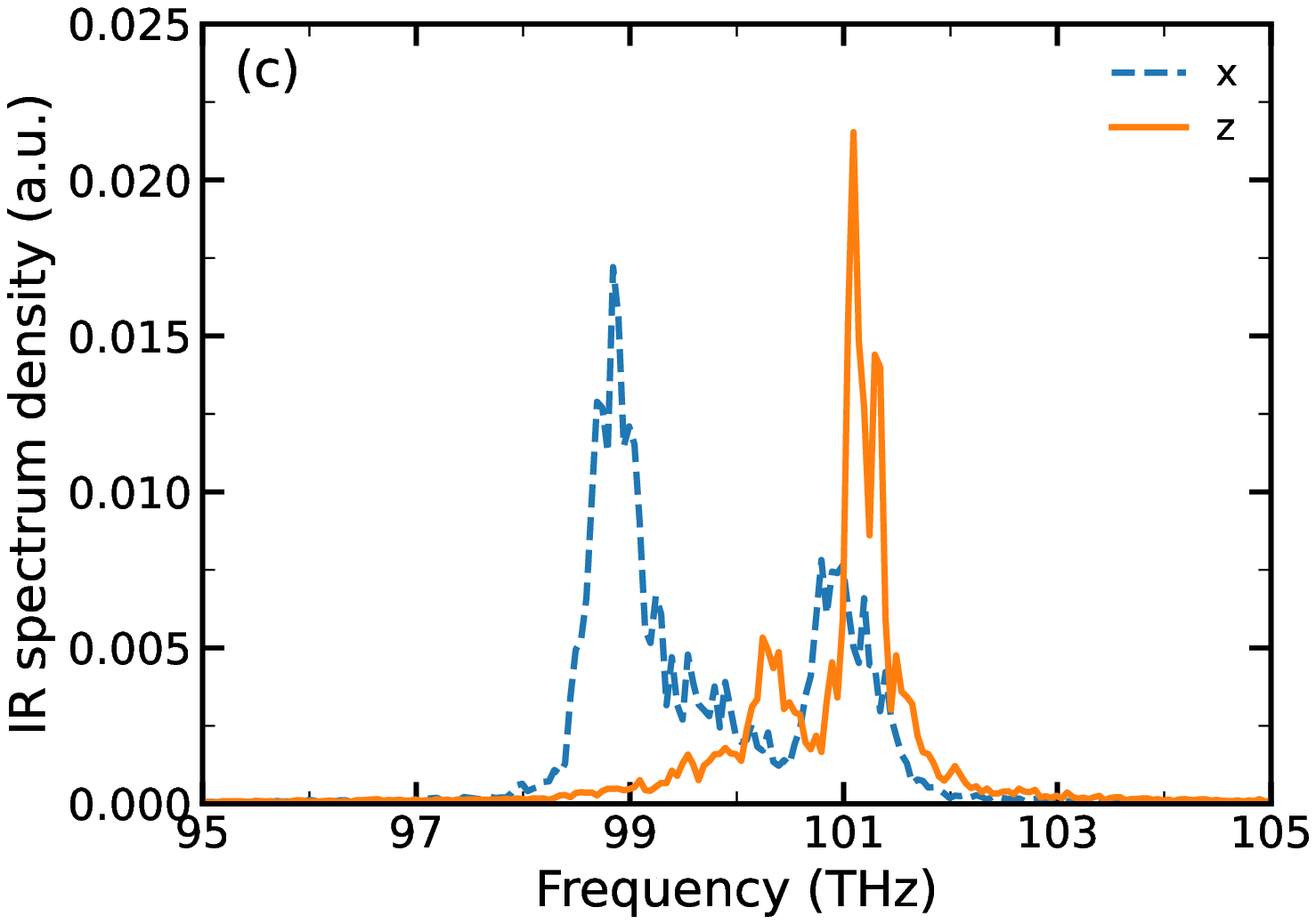}
\caption{(Color online) IR spectrum density along $x$ or $z$ axis for normal bulk water (a), polarized bulk water with SEF constraint (b) and surface water confined on a hexagonal lattice (c).} \label{F-spec}
\end{figure}

The IR spectrum density calculated from dipole autocorrelation function based on the linear response theory usually gives a good description about the electromagnetic field absorption processes. Fig.~\ref{F-spec} displays the IR spectrum density obtained from microcanonical trajectories in equilibrium simulations for unpolarized normal bulk water (a), polarized bulk water with SEF constraint (b) and surface water confined on a hexagonal lattice (c). Comparing Fig.~\ref{F-T-Freq}a and Fig.~\ref{F-spec}a, one can easily find that locations of the peaks perfectly coincide with each other. Moreover, the feature of slightly more efficient heating due to symmetric stretching mode near 99 $THz$ in Fig.~\ref{F-T-Freq}a is also captured in the corresponding spectrum density profile. Since normal bulk water is isotropic, the spectrum densities along $x$ and $z$ axis are almost the same, except for some statistical fluctuations.
For the polarized bulk water with pulses polarized parallel to the SEF along $z$ axis, the solid curve in Fig.~\ref{F-spec}b reveals a significant amplification of the spectrum at symmetric stretching frequency near 99 $THz$ and a large amount of reduction at asymmetric stretching frequency near 101 $THz$. The spectrum along $x$ axis is similar with the unpolarized case at asymmetric stretching area and a moderate reduction at symmetric stretching area.
Although the peaks' locations in these spectrum are in accord with the temperature jump in Fig.~\ref{F-T-Freq}a, the amplitude of enhancement in IR spectrum density along $z$ axis at the frequency of symmetric stretching for polarized bulk water is nearly four times of that corresponding to unpolarized bulk water, while the temperature jump is less than two times that of unpolarized case. Similarly, a moderate reduction is seen in the spectrum along $x$ axis at 99 $THz$, while there is only a small decrease in the corresponding temperature jump. This actually originates from the instantaneous change in orientational distributions of water molecules in dynamic response to the strong external laser pulse, which will be addressed in detail later. In the surface water system, the seesaw phenomenon of reduction and enhancement of symmetric and asymmetric stretching excitation due to the polarized pulse is well reproduced in the spectrum density. The temperature jump to IR spectrum density ratio at peaks in Fig.~\ref{F-T-Freq}-\ref{F-spec} is nearly the same for the surface water and the unpolarized bulk water.

As seen in the case of polarized bulk water, the IR spectrum density acquired from equilibrium simulations is insufficient to describe strong field absorption processes when the interaction between the system and the external field destroys the initial equilibrium configurations.
The absorption coefficient with regard to molecular excitation induced by a laser pulse is determined by the transition probability which is proportional to $cos^2\theta$ where $\theta$ is the angle between the transition dipole moment of a particular vibrational excitation and the direction of pulse's polarization. For unpolarized bulk water, where water dipoles are randomly distributed in all direction, the average of $cos^2\theta$ over all water molecules is always 1/3 no matter what the pulse's polarization direction is. However, anisotropic features will appear for polarized bulk water since a preferential direction exists. It is known that the transition dipole moment corresponding to excitation of the symmetric OH stretching is parallel to the direction of water dipole (i.e., axis of symmetry of water molecule), while that of the asymmetric stretching is vertical to water dipole. Thus, with pulses polarized along the water preferential direction the average $cos^2\theta$  for symmetric stretching is much larger than 1/3 due to a large population of small $\theta$. Meanwhile, $cos^2\theta$  for asymmetric stretching is less than 1/3 for the reduced population of water dipoles vertical to the pulse.
Shown in Fig.~\ref{F-ang} are the initial and final probability distributions of angle between the axis of symmetry of a water molecule and the polarization direction of a pulse  in polarized bulk water and surface water, which could depict a rough impression of the transition probability of symmetric stretching vibration near 99 $THz$ under radiations of pulses polarized along $z$ or $x$ axis. The initial distributions are the same as those from equilibrium simulations which are not shown here for simplicity. For the case in polarized bulk water, a large amount of water dipoles are initially located at $\theta\approx25^{\circ}$ to $z$ axis (solid orange curve) because of alignment to the applied SEF. The distribution changes significantly after irradiation of a 99 $THz$ pulse, becoming much less ordered during the absorption process of the pulse (dashed red curve). One can imagine that the actual transition probability of symmetric stretching vibration will be considerably smaller that predicted from the initial angular distribution. That is the reason why the temperature jump in polarized bulk water at 99 $THz$ in Fig.~\ref{F-T-Freq}a is much lower than the result predicted by the IR spectrum density in Fig.~\ref{F-spec}b when the pulse is polarized along $z$ axis. Analogously, the population reduction of water molecules at $90^\circ$ due to the absorption of pulses along $x$ axis enlarges the heating efficiency attributed to the excitation of the symmetric OH stretching vibration. The angular distribution in surface water is more or less the same before and after the irradiation due to the strong confinement exerted by the surface. In this case, the spectrum density thus remains eligible for giving a good description.

\section{Conclusion}
The prompt energy absorption from polarized MIR pulses in water systems through intramolecular resonant excitations are investigated using molecular dynamics simulation. It is found that the consequent heating based on the water's OH stretching vibrations at 99 and 101 $THz$ is very prompt, achieving over 100 $K$ temperature jump under the irradiation of a pulse with 1 $ps$ FWHM and 0.5 $V/nm$ maximum intensity, which is much more efficient than those based on intermolecular librational or translational vibrations. Bulk water polarized under the SEF and water monolayer confined on a hexagonal surface are explored here as two examples of water systems that possess orientational preference.
A general anisotropic phenomenon is manifested as a result of conditional resonant excitation of symmetric or asymmetric OH stretching vibration, depending on the relationships between the orientations of water molecules and the polarized direction of the pulses. In the case with the preferred dipole orientation of water molecule parallel to the polarized direction of the pulse, the energy absorption based on the symmetric stretching vibration near 99 $THz$ is enhanced dramatically, and the absorption according to the asymmetric stretching near 101 $THz$ becomes sizeably weakened. The results are reversed in the case with the pulses polarized vertical to the direction of the  preferential water dipoles, i.e., the asymmetric stretching absorption is amplified and the excitation of symmetric stretching is depressed. This seesaw characteristic of the efficiency shows the delicate relationships between the energy absorption processes and the specific situation of water molecules and the pulse polarization. On the other hand, water configurations experience significant changes during the dynamic processes of prompt and vast energy absorption, as suggested by the IR spectrum density. Since orientated water molecules are prevalent in chemical and biological systems, these findings emphasize the importance of orientation-dependent excitation of intramolecular vibrations which could significantly affect transitions and energy absorptions during relevant biochemical processes.


\section*{Acknowledgement}
The work was supported in part by the National Natural Science Foundation of China under Grant No. 11775049 and the China Postdoctoral Science Foundation under
 Grant No. 2021M690627.

\bibliography{ref-heat}

\end{document}